\documentclass[letterpaper]{article} % DO NOT CHANGE THIS
\usepackage{aaai2026}  % DO NOT CHANGE THIS
\usepackage{times}  % DO NOT CHANGE THIS
\usepackage{helvet}  % DO NOT CHANGE THIS
\usepackage{courier}  % DO NOT CHANGE THIS
\usepackage[hyphens]{url}  % DO NOT CHANGE THIS
\usepackage{graphicx} % DO NOT CHANGE THIS
\def\UrlFont{\rm}  % DO NOT CHANGE THIS
\usepackage{natbib}  % DO NOT CHANGE THIS AND DO NOT ADD ANY OPTIONS TO IT
\usepackage{caption} % DO NOT CHANGE THIS AND DO NOT ADD ANY OPTIONS TO IT
\usepackage{amsmath,amssymb}
\usepackage{booktabs}
\usepackage{tabularx}
\usepackage{array}
\usepackage{multirow}
\usepackage{xcolor}
\usepackage{nicematrix}
\usepackage{graphicx}
\usepackage{ragged2e}
\usepackage{colortbl}
\usepackage{tikz}
\usepackage{enumitem}
\usepackage[capitalise,noabbrev]{cleveref}
\crefname{section}{\S}{\S\S}
\usepackage{fvextra}

\DeclareUrlCommand\path{\def\UrlFont{\fontfamily{pcr}\selectfont}}

\newcommand{\promptlabel}[1]{%
  \par\addvspace{4pt}%
  \noindent{\itshape #1}%
  \par\nobreak\addvspace{1pt}}

\usetikzlibrary{positioning,arrows.meta,calc}
\definecolor{newblue}{RGB}{20,120,220}

\newcolumntype{L}{>{\RaggedRight\arraybackslash}X} 

\newcommand{\statTotalCases}{84}
\newcommand{\statJurisdictions}{11}
\newcommand{\statDomains}{13}
\newcommand{\statOfficialsN}{58}
\newcommand{\statOfficialsPct}{69\%}
\newcommand{\statThirdPartyN}{26}
\newcommand{\statThirdPartyPct}{31\%}
\newcommand{\statServicePct}{73\%}
\newcommand{\statNonServicePct}{27\%}
\newcommand{\statTopDomainOneName}{Justice \& Legal Services}
\newcommand{\statTopDomainOneN}{19}
\newcommand{\statTopDomainOnePct}{23\%}
\newcommand{\statTopDomainTwoName}{Regulatory Complaints}
\newcommand{\statTopDomainTwoN}{10}
\newcommand{\statTopDomainTwoPct}{12\%}
\newcommand{\statTopDomainThreeName}{Benefits \& Social Protection}
\newcommand{\statTopDomainThreeN}{9}
\newcommand{\statTopDomainThreePct}{11\%}
\newcommand{\statLLMTextPct}{87\%}
\newcommand{\statSeriesCases}{25}
\newcommand{\statQualitativeN}{76}
\newcommand{\statQualitativePct}{90\%}
\newcommand{\statQuantitativeN}{50}
\newcommand{\statQuantitativePct}{60\%}
\newcommand{\statBothTypesN}{42}
\newcommand{\statBothTypesPct}{50\%}
\newcommand{\statRespUseAiN}{21}
\newcommand{\statRespUseAiPct}{25\%}
\newcommand{\statRespRedesignServiceN}{13}
\newcommand{\statRespRedesignServicePct}{15\%}
\newcommand{\statRespSuppressDemandN}{14}
\newcommand{\statRespSuppressDemandPct}{17\%}

\definecolor{preprintbg}{RGB}{255,249,219}
\definecolor{preprintframe}{RGB}{201,162,39}
\newcommand{\preprintbox}[1]{%
  \par\addvspace{6pt}%
  \noindent
  \begin{tikzpicture}
    \node[
      draw=preprintframe,
      fill=preprintbg,
      line width=0.5pt,
      rounded corners=2pt,
      inner xsep=6pt,
      inner ysep=5pt,
      text width=\dimexpr\columnwidth-12pt-0.5pt\relax,
      align=justify
    ] {#1};
  \end{tikzpicture}%
  \par\addvspace{6pt}%
}

\begin{document} 

\title{Characterizing Agentic Flooding of Government Services} 
\author{
    Chris Schmitz\textsuperscript{\rm 1},
    Lewis Hammond\textsuperscript{\rm 2},
    Alan Chan\textsuperscript{\rm 3}
}
\affiliations{
    \textsuperscript{\rm 1}Centre for Digital Governance, Hertie School\\
    ch.schmitz@hertie-school.org \\ 
    \textsuperscript{\rm 2}Cooperative AI Foundation\\
    \textsuperscript{\rm 3}GovAI
}
\date{}

\maketitle

\begin{abstract}

AI agents are making it easier for the public to interact with government, such as by helping them apply for benefits, understand complex policies, and make their opinions heard. Although improving service accessibility is beneficial, any resulting surges in demand could strain unprepared government services.
We term such surges \textit{agentic flooding} of government services (``flooding'') and provide three contributions.
First, based on a collected dataset of \statTotalCases\ potential cases of flooding across \statJurisdictions\ jurisdictions, we posit that flooding is likely occurring widely today, mostly through large language models (LLMs) generating text cheaply. Second, we evaluate what services are most exposed to flooding. We develop a risk matrix to analyze a service's exposure, and suggest that near-term risk is highest for financially attractive, but complex services.
Finally, we map possible government responses to flooding.
Precedent suggests these responses will likely be sufficient to stop most cases of flooding, but the fastest to deploy -- friction-inducing measures like fees -- often trade off equitable access to public services.
Accordingly, we close by recommending near-term actions that may allow governments to mitigate flooding without invoking this trade-off.

\end{abstract}

\begin{links}
    \link{Dataset}{https://github.com/CLSchmitz/flooding-dataset}
\end{links}

\preprintbox{\small\textbf{Preprint.} This work will appear in the proceedings of the 9th AAAI Conference on AI, Ethics, and Society (AIES), October 12--14, 2026.}

\section{Introduction}
\label{sec:introduction}

\begin{figure}[t!]
\centering
\includegraphics[width=\columnwidth]{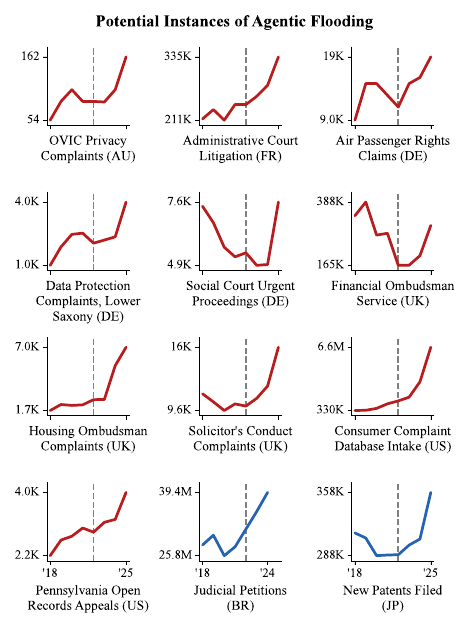}
\caption{Annualized submission volumes (2018--2025) for 12 government services. For each, either government officials (red) or reputable secondary sources (blue) have asserted AI involvement in surges. Only some surges begin around the 2022 release of ChatGPT (dashed lines).}%
\label{fig:sparklines}
\end{figure} 

\begin{table*}[t]
\centering
\small
\renewcommand{\arraystretch}{1.4}
\caption{Types of ``administrative burden'' citizens face when interacting with government \citep{moynihanAdministrativeBurdenLearning2015}, the AI agent capabilities that may reduce them, and technologies that enable these capabilities.}
\label{tab:frictions}
\begin{tabularx}{\textwidth}{@{} >{\RaggedRight\arraybackslash}p{2cm} >{\RaggedRight\arraybackslash}p{3.8cm} L L @{}}
\toprule
\textbf{Burden Type} & \textbf{Specific Burden} & \textbf{Agent Capability} & \textbf{Enabling Technologies} \\
\midrule
\multirow{3}{2cm}{\textbf{Learning Costs}}
  & Understanding complex rules and procedures
  & Summarizing complex regulations and requirements in plain language
  & LLMs with large context \\
  \arrayrulecolor{lightgray}\cmidrule(l){2-4}\arrayrulecolor{black}
  & Uncertainty about eligibility
  & Evaluating personal data against policy and entitlement rules
  & Context augmentation, directly or, e.g., via Model Context Protocol (MCP) \\
  \arrayrulecolor{lightgray}\cmidrule(l){2-4}\arrayrulecolor{black}
  & Unawareness of services or entitlements
  & Scanning or continuously monitoring for entitlements; proactive alerts
  & Web search tools; pipelines with scheduled execution \\
\midrule
\multirow{4}{2cm}{\textbf{Compliance Costs}}
  & Legal writing; filling forms
  & Context-specific, sophisticated writing
  & LLMs with user context\\
  \arrayrulecolor{lightgray}\cmidrule(l){2-4}\arrayrulecolor{black}
  & Confusing government websites and service portals
  & Navigating and interacting with websites autonomously
  & Browser use tools; vision-language models \\
  \arrayrulecolor{lightgray}\cmidrule(l){2-4}\arrayrulecolor{black}
  & Locating and assembling supporting documents
  & Retrieving, formatting, and attaching required supporting materials
  & File system access; multimodal document parsing \\
  \arrayrulecolor{lightgray}\cmidrule(l){2-4}\arrayrulecolor{black}
  & Managing response deadlines and follow-ups
  & Monitoring deadlines; alerting users or submitting replies autonomously
  & Asynchronous task queues; calendar API integration \\
\midrule
\multirow{3}{2cm}{\textbf{Psychological Costs}}
  & Having to re-explain case context to many contacts
  & Maintaining full case context across multi-step processes
  & Memory; context management \\
  \arrayrulecolor{lightgray}\cmidrule(l){2-4}\arrayrulecolor{black}
  & Uncertainty about case progress and next steps
  & Polling for case updates and notifying the user only when action is required
  & Scheduled execution; access to web or API endpoints \\
  \arrayrulecolor{lightgray}\cmidrule(l){2-4}\arrayrulecolor{black}
  & Repeated dehumanizing bureaucratic interactions
  & Handling rejections, follow-up calls, and status chasing autonomously
  & Tools for external communication, e.g. email and voice synthesis APIs \\
\bottomrule
\end{tabularx}
\end{table*}

AI agents increasingly help the public interact with government. For example, they can already generate legally coherent text for an official complaint, or parse complex policy documents into plain language. \citet{iscenkoGooglesAIEconomy2026} classify about 1\% of requests to Google Gemini as helping with government interaction.
These capabilities reduce submission costs for citizens and make government more accessible, a desirable public value \citep{mooreCreatingPublicValue1995}.

However, such reductions in cost could also cause surges in the complexity or quantity of requests. 
Evidence of such surges is emerging: the Australian government has considered reintroducing fees for Freedom-of-Information (FOI) requests citing a wave of AI-generated submissions \citep{canalesAddictionSecrecyOpposition2025}; some German social courts largely attribute a 55\% year-on-year rise in caseload in 2025 to AI-generated claims \citep{LTO2026Sozialgerichte}; \citet{shahAccessJusticeAge2026} find a ``dramatic increase'' in both the share and average length of self-represented cases before US federal courts since 2022, presumably driven by AI-aided citizen self-representation.

Key questions about such surges remain open. In particular, we need to know how widespread they are, how serious a risk they pose, and what governments can do about them. Given the rapid pace of AI capabilities advancements, there is a clear and pressing need for greater understanding.

This work provides a structured treatment of the phenomenon we term \textit{agentic flooding} of government services (or ``flooding'' hereafter): surges in the volume or complexity of requests they receive, which (1) strain their capacity and (2) are enabled by agents reducing the cost of interacting with such services (\cref{sec:service-flooding}).  We address three questions:

\paragraph{Is flooding happening -- and if so, in what forms? (\cref{sec:exists})} We compile a dataset (\cref{sec:approach}) of \statTotalCases\ recent surges in service demand, each of which officials or reputable secondary sources attribute to AI use (\Cref{fig:sparklines}).
The data is consistent with agentic flooding occurring widely today, though our methodology does not permit causal or quantitative claims about AI's impact. 
Almost all the cases we identify share a mechanism: LLMs generating large amounts of text, paired with humans navigating the rest of the process manually. More sophisticated use of agents, such as autonomous website navigation, is not yet evident. 

\paragraph{How big is the risk? (\cref{sec:severity})} The risk of flooding depends on the progress and diffusion of AI capabilities, and varies significantly across countries and services.
Analyzing our data, we posit that the near-term risk of flooding is highest for financially attractive services where complex submission requirements have historically suppressed demand, such as court claims and tax returns. These properties can be assessed in services now, and such assessments can be updated as new information about AI capabilities arrives. We construct a risk matrix to guide more detailed assessment of individual services. %

\paragraph{How could governments respond? (\cref{sec:responses})} We map government response options, grouped under two strategies: suppressing demand (e.g. fees, rate limits, in-person requirements) and increasing capacity (e.g. staffing more people, deploying AI, or redesigning the service interface). Precedent suggests both approaches are sufficient to address flooding, but demand suppression is far faster to deploy. However, it decreases the quality and accessibility of government services, and can introduce procedural inequality.

Preparing proactively could help governments avoid this trade-off. To that end, we recommend three near-term actions (\cref{sec:recommendations}): auditing service vulnerability, integrating digital identity into the most exposed services, and clarifying the legality of resilience-building measures.

\section{Agentic Flooding of Government Services}
\label{sec:service-flooding} 

This section defines \textit{agentic flooding} of government services (\cref{sec:definition}) and analyzes its potential causes (\cref{sec:ai-burdens}).

\subsection{Definitions}
\label{sec:definition}

Throughout this work, we use \textit{AI agents} (or ``agents'') to mean LLM-based systems with some degree of autonomy and tool access \citep{kasirzadehCharacterizingAIAgents2025}.
Our definition includes current-generation consumer-facing LLMs, as they typically integrate tools such as web search \citep{sadeddineLargeLanguageModels2025}. 

We use \textit{government services} as a deliberately broad term, to mean both public services -- such as welfare or infrastructure -- and non-service channels of government interaction, e.g. public participation, judicial procedures, or freedom of information requests \citep{osbornePublicServicedominantLogic2018}.

We use \textit{the public} and \textit{users} interchangeably to refer to all individuals interacting with government services.

With these terms, we can define agentic flooding. 

\begin{center}
\begin{tikzpicture}
\node[draw=gray, fill=newblue!10, rounded corners=3pt, inner sep=10pt, text width=0.9\columnwidth] {
    \textit{Agentic flooding} of a government service is a surge in the volume or complexity of requests it receives, which \\[1ex]
    \textit{(i)} is caused by AI agents interacting with it; and \\[1ex]
    \textit{(ii)} substantially strains the service.
};
\end{tikzpicture}
\end{center}

\label{sec:variations}

We distinguish between two types of flooding: increases in the number of requests (\textit{quantitative} flooding), and increases in the complexity of each request (\textit{qualitative} flooding). For example, agents that help users discover eligible services may increase request volume, whereas agents that draft text may lengthen the average complaint letter. An instance of flooding could be both qualitative and quantitative. For example, LLM hallucinations could result in high volumes of complex, but incorrect, applications. Both of these effects, which \citet{blundellExtensiveIntensiveMargins2013}  term the ``extensive'' and ``intensive'' margins of work, increase processing burden. However, some potential responses only address one type (\cref{sec:responses}). For example, introducing rate limits is unlikely to ease qualitative flooding.

\subsection{Potential Causes of Flooding}
\label{sec:ai-burdens}

AI agents could cause a surge in the volume or complexity of requests by (1) reducing the costs of interacting with the government service or (2) increasing the benefits of doing so. As such, both users and middle-man organizations, e.g. law firms, have incentives to deploy agents to interact with government services. Any resulting increases in demand could strain the service because many services are only designed to handle a limited amount of demand.

\paragraph{Agentic Capabilities Could Reduce Costs.}
AI agents could reduce \textit{administrative burdens}: the costs of participating in bureaucratic procedure which users face when interacting with government \citep{hallingAdministrativeBurdenCitizen2024}. \citet{moynihanAdministrativeBurdenLearning2015} group these costs into three types: learning, compliance, and psychological costs.

In \Cref{tab:frictions}, we map how agent capabilities could reduce all three costs. \textit{Learning costs}, the cognitive effort of understanding government, are addressed most directly by agents' ability to parse long contexts and personalize communication to users \citep{kasneciChatGPTGoodOpportunities2023}. \textit{Compliance costs}, the practical effort of completing required steps, fall as agents grow more capable of high-quality digital work \citep{drouinWorkArenaHowCapable2024}. 
\textit{Psychological costs}, the emotional toll of repeated bureaucratic contact, drop as agents handle more of that contact.

Reductions in administrative burdens often raise demand for services. For example, the introduction of an app to request municipal services in Boston led to a 33\% increase in reports \citep{obrienUnchartedTerritorialityCoproduction2016}, and online voter registration options consistently increase enrolment numbers \citep{garnettRegistrationInnovationImpact2022}. 

At a conceptual level, rational-actor models propose that users submit requests when the benefit they expect exceeds the cost of submission. \citep{nicholsTargetingTransfersRestrictions1982, zeckhauserStrategicSortingRole2019}. 
Individual users rarely act financially optimally -- their behavior is shaped by many other factors, such as psychological frictions or stigma \citep{currieTakeSocialBenefits2004, bhargavaPsychologicalFrictionsIncomplete2015}. But the rational-actor model often accurately describes aggregate behavior across a population \citep{vanoorschotNonTakeUpSocialSecurity1991}. We would therefore expect requests to increase when AI agents decrease submission costs. 

\paragraph{Agentic Capabilities Could Increase Benefits.}
Agentic capabilities could also increase the benefits of interacting with a service. For example, agents could rewrite Request for Information (RFI) responses to be more convincing, increasing the likelihood that the responses' views are reflected in policy. AI agents could also cheaply help to ensure that all information relevant to an application is gathered, ensuring that applicants receive the maximal benefit. Analogously, tax accountants help to ensure that clients receive their maximum possible tax refund. 

\paragraph{Incentives to Deploy Agentic Capabilities.}
Beyond the user-level incentives, there are strong commercial incentives to deploy agentic capabilities in ways that increase demand for government services. The business model is well-established: middle-man organizations offer to handle complex government interactions on behalf of users in exchange for a cut of the earned entitlements, growing the number of claims to government.
The most prominent example is claims management companies, which have long mass-submitted benefit and compensation claims through templated submissions, such as for flight delay compensation and UK personal injury claims.
Companies advertising their use of AI to similar effect already exist \citep{TimeMoneyDoNotPay2026}.

Further, adversarial actors may explicitly seek to flood services. For example, they may aim to destabilize critical infrastructure \citep{piedrahitaAIPosesRisks2026}, or to incapacitate agencies. Even good-faith users may be incentivized to abuse the system. Analoguously, ``vexatious litigants'' file large numbers of meritless court claims, each of which they are entitled to file, but which collectively overwhelm the court \citep{rustVexatiousLitigantProblem2024}. Thus far, transaction costs have kept such cases rare enough for ad-hoc management; agent capabilities could make these cases much more frequent.

\paragraph{Increased Demand Can Cause Strain.} 
Increased demand for government services is not in itself harmful, but agencies often cannot meet it. Many services run close to operational capacity because of budget constraints, operational inefficiency, or legal obligations to minimize overhead \citep{7BHOEinzelnorm}. 
Some governments know of and tolerate, or even deliberately maintain, administrative burdens which suppress take-up \citep{peetersAdministrativeExclusionInfrastructurelevel2023}. 

Capacity strain can have both direct operational consequences -- like growing backlogs or missed response deadlines -- and broader fiscal and policy effects, such as requiring budget adjustments (\cref{sec:severity}).
Collectively, these effects may erode user experience of public services, which strongly predicts trust in government \citep{vandewallePublicServicePerformance2003,christensenTRUSTGOVERNMENTRelative2005}.

\section{Evidence Collection}
\label{sec:approach}

To compile a dataset of real-life cases where flooding may be occurring, we scan government services in 12 countries (\cref{sec:scope}).
Because AI involvement in demand surges is difficult to prove, we only include cases where a government official or credible third party has asserted such involvement (\cref{sec:case-selection}). This enables a broad scan but trades off statistical representativeness. To collect cases, we employ an LLM-aided qualitative coding workflow (\cref{sec:methodology}).

\subsection{Scope}
\label{sec:scope}

We scan for publicly accessible information about government services in 12 countries: Australia, Brazil, Denmark, Estonia, France, Germany, Japan, the Netherlands, Singapore, South Korea, the United Kingdom, and the United States. We choose countries where it is common to report publicly on administrative matters, aiming for variation in geography and in two core dimensions of public administration scholarship: administrative tradition  \citep{painterTraditionPublicAdministration2010} and digital government maturity \citep{oecdGovernmentGlance20252025}.

In each of these countries, we scan the same fixed list of \statDomains\ government domains. We set this list by combining public-facing lists of government services from three studied countries (UK, US, and France). Ten domains cover public services, e.g. benefits and social protection,  citizenship, or jobs and pensions. Three cover government processes that are not public services, but in which flooding may also occur: participatory processes; regulatory complaints and reporting; and transparency and access to information. 
Appendix~\ref{app:scope} 
describes the detailed methodology with which we set both scopes.

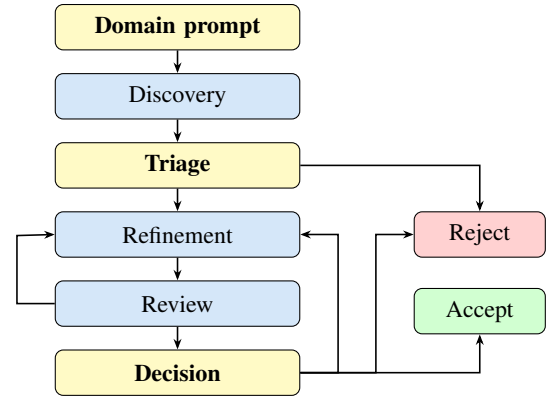
\begin{figure}
\centering
\begin{tikzpicture}[
  node distance=0.3cm,
  llm/.style={draw, rectangle, rounded corners=3pt, fill=newblue!18,
    text width=3cm, minimum height=0.6cm, align=center, font=\small},
  human/.style={draw, rectangle, rounded corners=3pt, fill=yellow!28,
    text width=3cm, minimum height=0.6cm, align=center, font=\small\bfseries},
  term/.style={draw, rectangle, rounded corners=3pt,
    text width=1.5cm, minimum height=0.55cm, align=center, font=\small},
  arr/.style={-{Stealth[length=4pt,width=3pt]}, semithick}
]
\node[human]                      (prompt) {Domain prompt};
\node[llm,   below=of prompt]     (disc)   {Discovery};
\node[human, below=of disc]       (triage) {Triage};
\node[llm,   below=of triage]     (iter)   {Refinement};
\node[llm,   below=of iter]       (rev)    {Review};
\node[human, below=of rev]        (hdec)   {Decision};
\node[term, fill=red!18,  right=1.5cm of iter, minimum height=0.6cm] (dep)   {Reject};
\node[term, fill=green!18, below=0.4cm of dep, minimum height=0.6cm]  (final) {Accept};
\draw[arr] (prompt) -- (disc);
\draw[arr] (disc)   -- (triage);
\draw[arr] (triage) -- (iter);
\draw[arr] (iter)   -- (rev);
\draw[arr] (rev)    -- (hdec);
\draw[arr] (triage.east) -| (dep.north);
\draw[arr] (hdec.east) -| (final.south);
\draw[arr] (hdec.east)
  -- ($(dep.west |- hdec)+(-0.5,0)$)
  -- ($(dep.west)+(-0.5,0)$)
  -- (dep.west);
\draw[arr] (rev.west) -- ++(-0.5,0) -- ++(0,0.9) -- (iter.west);
\draw[arr] (hdec.east) -- ++(0.5,0) |- (iter.east);
\end{tikzpicture} 
\caption{Overview of LLM-aided pipeline for collection of cases. Yellow: human step. Blue: LLM step.}
\label{fig:pipeline-overview}
\end{figure}

\begin{table*}[t]
\centering
\small
\renewcommand{\arraystretch}{1.2}
\caption{Ten illustrative cases of \textit{agentic flooding} in our dataset.}
\label{tab:exemplary-cases}
\begin{tabularx}{\textwidth}{@{} lllXl @{}}
\toprule
\textbf{ID} & \textbf{Jurisdiction} & \textbf{Service Domain} & \textbf{Service Name} & \textbf{Flooding Type} \\
\midrule
    A & Australia & FOI & Federal FOI Requests  & Quantitative \\
    B & Germany & Justice \& Legal & Social Court Lawsuits &  Qual.\ \& Quant. \\
    C & Germany & Public Participation & Environmental Impact Assessment (UVP) Public Comments & Qual.\ \& Quant. \\
    D & Brazil & Employment & Temporary Disability Benefit (INSS Auxílio-doença) & Qualitative \\
    E & United Kingdom & Justice \& Legal & Money Claims Online (MCOL) / Online Civil Money Claims & Qual.\ \& Quant. \\
    F & United Kingdom & Public Participation & Local Plan Consultations (Regulation 18 and 19) & Qual.\ \& Quant. \\
    G & Japan & Public Participation & Strategic Energy Plan Public Consultations & Qual.\ \& Quant. \\
    H & South Korea & Justice \& Legal & Civil E-Litigation (Electronic Payment Orders) & Qualitative \\
    I & Netherlands & Justice \& Legal & Municipal WOZ Valuation Objections & Qual.\ \& Quant. \\
    J & United States & Citizenship & Voter Challenge and Voter Roll Purge Submissions & Qual.\ \& Quant. \\
\bottomrule
\end{tabularx}
\end{table*}

\subsection{Case Selection and Attribution}
\label{sec:case-selection}
 
Demand for public services is shaped by many interrelated factors including policy changes, spurious events, and broader social and economic developments; processing indicators like backlog sizes and wait times are similarly confounded \citep{oecdWaitingTimePolicies2013}. 
Deeper analysis may mitigate these problems in individual cases, especially where submission contents are public \cite{shahAccessJusticeAge2026}, but it is not (yet) feasible at scale.

As a way to partially address this issue, we select cases conservatively. 
In addition to evidence of the two criteria in our definition, we require that the affected government body itself -- or a reputable, relevant source -- attribute a change in demand patterns to AI use by the public.
We pose three separate inclusion criteria per case:

\begin{enumerate}
    \item \textbf{A plausible, specific mechanism} by which AI could reduce transaction costs for the service -- e.g. that LLMs help write a complaint which can then be submitted online. We judge plausibility ourselves based on the design of the service.
    \item \textbf{Evidence of change} in demand patterns for the service that is consistent with what the mechanism in (1) could cause, found in official statements or actions (or another high-quality source), and reflected in  primary (self-collected) volume data if available. Trends in the volume data alone do not fulfill the criterion.
    \item \textbf{External attribution} of that change to AI use, by either (a) a government source -- e.g. a press statement, official communication, or policy update -- or (b) a reputable secondary source -- e.g. an established news outlet, professional legal or administrative publication, or peer-reviewed work.
\end{enumerate}

Our methodology produces a broad dataset but does not afford causal, quantitative, or statistical claims. We further discuss limitations in \cref{sec:limitations}.

\subsection{Methodology}
\label{sec:methodology}

\paragraph{Data Format.} We code all cases into a standardized template, provided in the dataset repository and outlined in Appendix~\ref{app:pipeline}. 
All figures for which we provide statistics are captured in a structured format in this schema. We do not perform additional analysis on free-text fields, except to highlight illustrative examples. For each case, we attempt to collect:

\begin{itemize}
    \item Free-text descriptions of the service and its submission interface, the mechanism by which flooding may occur, and any impact and government response observed.
    \item Structured fields classifying the case type and mechanism along the typologies in \cref{sec:ai-burdens,sec:variations}, whether it meets each of the three inclusion criteria, and who asserts AI involvement in the change in requests.
    \item Annualized volume statistics from 2018 to 2025 in a case-specific unit (e.g.\ requests filed, backlog size).
    \item Metadata, e.g.\ country, government body, and domain; and links to all sources used in compiling the case.
\end{itemize}

\paragraph{Case Collection Pipeline.}

We employ an LLM-aided research and qualitative coding workflow \citep{yeHybridSemiAutomatedWorkflow2024}. As outlined in \Cref{fig:pipeline-overview}, the pipeline has three stages. During \textit{discovery}, a country-domain combination (e.g. ``France, Health Services'') is scanned for candidate services. In \textit{iteration}, each case cycles between LLM calls to improve case data and review case quality. In \textit{finalization}, a human reviewer validates the case, makes required corrections, and chooses whether to include it in the dataset. Detailed prompts and scaffolding are listed in Appendix~\ref{app:pipeline}.

\paragraph{Preventing Hallucinations.} Our approach contains three safeguards against the risk of LLM hallucinations \citep{huangSurveyHallucinationLarge2025}. First, we perform human verification of all cases, before both iteration and finalization. Second, we force structured responses from all LLM calls, and maintain each case in a standardized, JSON-based data format. This format enables us to verify key fields, such as the names of government agencies and numeric case counts, deterministically; it also allows metadata, such as iteration count and review comments, to be tracked and passed to each LLM call.
Third, we extract accessed web URLs deterministically from the metadata of LLM API responses, rather than requiring the models themselves to generate them as part of the response. This guarantees all cited sources are real, human-verifiable websites, which we visit to confirm key figures.

\section{Is Agentic Flooding Happening?}
\label{sec:exists}

The collected data is consistent with flooding occurring today, across a wide range of government domains and jurisdictions. Almost all cases occur through one mechanism: LLM-generated text submitted by users who navigate the rest of the process manually. 

\paragraph{Dataset Overview.}

We document \statTotalCases\ cases across \statJurisdictions\ jurisdictions and \statDomains\ service domains. Full case data are available in the linked dataset repository. \Cref{tab:exemplary-cases} contains ten illustrative cases to which the subsequent analysis refers.

These \statTotalCases\ cases are the product of a broad scan followed by strict inclusion criteria: of the 2288 candidate services named during discovery -- roughly 190 per country -- fewer than one in twenty met all three criteria in \cref{sec:case-selection}. The biggest constraint on inclusion was requiring official or third-party attribution of the surge to AI.

Of the \statTotalCases\ cases, government officials assert AI involvement (with or without third-party corroboration) in \statOfficialsN\ (\statOfficialsPct); solely third-party sources assert it in the remaining \statThirdPartyN\ (\statThirdPartyPct). Public and judicial services are more represented than non-service interaction channels, such as participation procedures and transparency requests (\statServicePct\ and \statNonServicePct\ of cases, respectively).
The most frequently represented domains are \statTopDomainOneName\ (\statTopDomainOnePct, N=\statTopDomainOneN), followed by \statTopDomainTwoName\ (\statTopDomainTwoPct, N=\statTopDomainTwoN) and \statTopDomainThreeName\ (\statTopDomainThreePct, N=\statTopDomainThreeN).

\paragraph{Strength of Evidence.} The dataset provides evidence that AI use is increasing the volume and complexity of requests for at least some government services. The cases span jurisdictions, service types, and reporting outlets; in most cases, the affected government body itself asserts AI involvement. Any alternative explanation would have to account for independent misattribution across all of these criteria.

However, our data-gathering remains exploratory and diagnostic, and affords no causal or quantitative claims about either (a) the prevalence of flooding or (b) the role of agents. Demand is driven by many factors, 
and some volume series in \Cref{fig:sparklines} begin rising before the release of ChatGPT. Moreover, because we require explicit public attribution, the dataset likely undercounts the cases that meet our definition of flooding (\cref{sec:definition}). \Cref{sec:limitations} discusses these limitations.

\paragraph{Capabilities Enabling Flooding.}
In most of our cases (\statLLMTextPct), flooding occurs when LLMs help to cheaply generate legally sophisticated text. Most of the services in the dataset accept text, often of arbitrary length, e.g. FOI requests [Case A], planning consultations [Case F], or public comments [Case C]. 

There are two potential explanations for this pattern. First, text generation is more mature and accessible than agentic capabilities such as browser navigation \cite{zhouWebArenaRealisticWeb2024}.
Second, sampling bias: cases largely enter our dataset because officials or domain experts noticed an anomalous change in the pattern of submissions (\statOfficialsPct, N=\statOfficialsN). LLM-generated text can be identified by the content of the submission itself, whereas many other types of agent-aided submissions may look identical to human ones if the submission format is constrained -- e.g. by short fields or structured online forms. Any effect such submissions have on request patterns could only be observed via the quantity of requests, which is also affected by various other factors \citep{oecdWaitingTimePolicies2013}. It is plausible that officials are thus far attributing such quantitative changes to AI less frequently.

\paragraph{Types of Flooding Observed.}

We code \statQuantitativePct\ (N=\statQuantitativeN) of cases as quantitative flooding and \statQualitativePct\ (N=\statQualitativeN) as qualitative, with \statBothTypesPct\ (N=\statBothTypesN) exhibiting both types. These numbers are largely consistent with the above pattern of LLM text generation, which both allows \textit{more} users to submit requests to free-text service interfaces -- e.g. the less legally literate submitting litigation [Case H] -- and makes these requests \textit{longer}, as in [Case B], where some letters spanned over 4000 pages.

We observe cases of both adversarial and non-adversarial flooding in the dataset. While most cases show ordinary users acting in good faith, some involve individuals acting in bad faith for personal gain, e.g. attempting to get disability benefits with AI-generated medical certificates [Case D]. A few involve organized adversarial campaigns, notably mass-submitting FOI requests or voter roll challenges [Case J].

\section{How Big Is the Risk?}
\label{sec:severity}

\begin{figure}[t!]
\centering
\begingroup
\providecommand{\riskup}{\textcolor{red!75!black}{$\uparrow$}}
\providecommand{\riskdown}{\textcolor{green!50!black}{$\downarrow$}}

\setlength{\fboxsep}{7pt}
\setlength{\fboxrule}{0.45pt}
\centering
\fcolorbox{gray!25}{gray!4}{%
\begin{minipage}{\dimexpr\columnwidth-2\fboxsep-2\fboxrule-2pt\relax}
\footnotesize
\raggedright
\dimen6=\linewidth %
\settowidth{\dimen2}{14}%
\settowidth{\dimen3}{\riskup}%
\settowidth{\dimen4}{\riskdown}%
\ifdim\dimen4>\dimen3 \dimen3=\dimen4\fi
\advance\dimen2 by \dimen3
\advance\dimen2 by 12pt %
\noindent\hbox{\hspace{\dimen2}%
\begin{tikzpicture}[x=0.58cm, y=0.58cm, font=\footnotesize]
  \foreach \x in {0,...,2} {
    \foreach \y in {0,...,2} {
      \pgfmathtruncatemacro{\band}{max(\x,\y)}
      \ifnum\band=2
        \def\cellcolor{red!18}
      \else\ifnum\band=1
        \def\cellcolor{yellow!28}
      \else
        \def\cellcolor{green!18}
      \fi\fi
      \fill[\cellcolor] (\x,\y) rectangle ++(1,1);
    }
  }
  \draw[draw=gray!55, line width=0.28pt] (0,0) rectangle (3,3);
  \foreach \n in {1,2} {
    \draw[draw=gray!55, line width=0.22pt] (\n,0) -- (\n,3);
    \draw[draw=gray!55, line width=0.22pt] (0,\n) -- (3,\n);
  }
  \node at (0.5, 2.5) {\textbf{A}};
  \node at (1.5, 1.5) {\textbf{B}};
  \node at (2.5, 0.5) {\textbf{C}};
  \draw[-{Stealth[length=2.1mm]}, line width=0.45pt, gray!60] (0,-0.18) -- (3.18,-0.18);
  \node[anchor=north east, text=gray!70, inner sep=0, yshift=-4pt] at (3.18,-0.18) {Severity};
  \draw[-{Stealth[length=2.1mm]}, line width=0.45pt, gray!60] (-0.18,0) -- (-0.18,3.18);
  \node[anchor=south west, text=gray!70, inner sep=0, yshift=4pt] at (-0.18,3.18) {Likelihood};
\node[anchor=west, inner sep=6pt] at (3.3, 1.5) {%
    \setlength{\fboxsep}{4pt}\setlength{\fboxrule}{0.3pt}%
    \dimen7=\dimexpr\dimen6-\cmidrulekern-\dimen2-2.1cm-6pt-2\fboxsep-2\fboxrule-3pt\relax
    \fcolorbox{gray!55}{white}{%
      \vbox to \dimexpr1.74cm-2\fboxsep-2\fboxrule\relax{%
      \vfil
      \hbox to \dimen7{%
      \renewcommand{\arraystretch}{1.0}%
      \begin{tabular}{@{}l@{\hspace{4pt}}l@{}}
        & \itshape\textcolor{gray!70!black}{Sample services} \\
        \textbf{A} & Public comment procedures \\
        \textbf{B} & Benefits appeals \\
        \textbf{C} & Emergency services
      \end{tabular}\hfil}%
      \vfil}}};
\end{tikzpicture}}\par
\vspace{10pt} 
\renewcommand{\arraystretch}{1.15}
\begin{tabularx}{\linewidth}{@{} r @{\hspace{5pt}} c @{\hspace{7pt}} >{\raggedright\arraybackslash}X @{}}
  & & \bfseries Likelihood \\
  \cmidrule(r){3-3}
  \addlinespace[2.5pt]
  & & \itshape Agent Capabilities and Access \\
  \addlinespace[2.5pt]
  1  & \riskup & Maturity of agent capabilities \\
  2  & \riskdown & Cost to access the required capabilities \\
  \addlinespace[5pt]
  & & \itshape Service Features \\
  \addlinespace[2.5pt]
  3  & \riskup   & Susceptibility of service interface \\
  4  & \riskup & Effort required to submit \\
  5  & \riskup   & Expected benefit of a successful submission \\
  \addlinespace[8pt]
  & & \bfseries Severity \\
  \cmidrule(r){3-3}
  \addlinespace[2.5pt]
  & & \itshape Operational Impact \\
  \addlinespace[2.5pt]
  6  & \riskup   & Effort required to process each submission \\
  7  & \riskdown & Efficiency of identity verification \\
  8  & \riskdown & Ease of scaling processing capacity \\
  9 & \riskup   & Current utilisation of capacity \\
  \addlinespace[5pt]
  & & \itshape Response Impact \\
  \addlinespace[2.5pt]
  10 & \riskup & Rigidity of budget and entitlement policy \\
  11 & \riskup & Processing effort the law mandates per request \\
  12 & \riskup & Gap between assumed and full take-up \\
  13 & \riskup & Downstream cost of a successful submission \\
\end{tabularx}
\end{minipage}%
}
\endgroup
\caption{Risk matrix of factors which may predict the likelihood and severity of \textit{agentic flooding} for a given government service, with three illustrative services mapped. Arrows indicate whether a higher value raises ($\uparrow$) or lowers ($\downarrow$) risk.}
\label{fig:risk-matrix}
\end{figure}
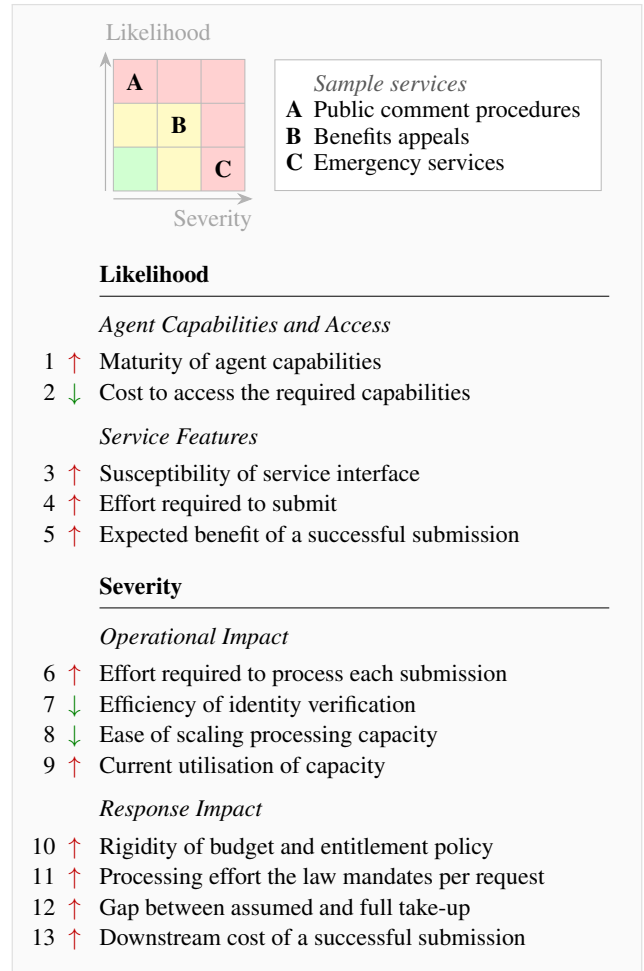

The risks posed by flooding depend on how quickly AI capabilities progress and diffuse, and vary with the service and jurisdiction affected.
A key question is therefore which services are most at risk. 
Analyzing our dataset, we posit that near-term risk is highest for financially lucrative services with complex applications and legal processing obligations, such as tax returns (\cref{sec:where-risk}).
We propose a risk matrix (\cref{sec:risk-matrix}) to more rigorously evaluate the likelihood and severity of a specific service being flooded.

\subsection{Which Services Are Most Vulnerable?}
\label{sec:where-risk}

The services in our dataset share several commonalities. Most obviously, they overwhelmingly accept free-form text through open digital channels, and the agent capability required to flood them -- LLM text generation -- is freely accessible (\cref{sec:exists}). 

The observed impacts of flooding so far are moderate. While we find evidence of overworked public servants, calls for additional budget, and calls to reform legal response obligations [Case E], the strain is modest compared to precedent, e.g. from mass comment campaigns \citep{ballaRespondingMassComputerGenerated2022}. Where governments respond, e.g. by re-introducing FOI fees [Case A] and batch-dismissing consultation responses [Cases C, G], these responses also appear effective. 

There are some higher-severity cases in our dataset, e.g. objections to tax valuations [Case I], social court lawsuits [Case B], and civil claims [Cases E, H]. These share two properties. First, each successful submission is individually consequential: it is financially valuable to the claimant if successful, or triggers legally mandated processing effort. Second, each service's resilience has historically been owed to friction rather than design. With few structural limits on submission length or volume, the complexity of submitting -- and the legal or professional knowledge it demanded -- historically gated who submitted, and how well. 

Given our inclusion criteria (\cref{sec:approach}), these patterns may also reflect which cases officials notice and report. Nonetheless the risk of flooding appears highest for services with these two properties: services that are financially attractive, and whose demand has so far been suppressed by friction or tacit knowledge. These could include tax administration and court systems. Notably, an initial assessment of services against these two properties does not require forecasting AI progress. They can be analyzed today, as we discuss in \cref{sec:recommendations}.

\subsection{Risk Matrix for Flooding}
\label{sec:risk-matrix}

\Cref{fig:risk-matrix} describes a risk matrix \citep{kaplanQuantitativeDefinitionRisk1981} for the flooding of a given service, containing factors that may predict the \textit{likelihood} that a service is flooded and the \textit{severity} if it is. We construct this matrix by combining our data with theory and precedent from past demand surges \citep{ballaRespondingMassComputerGenerated2022, verhoevenMyHowHave2024, woodAgencyPerformanceChallenges2017}. Such a matrix could be used to map a government's exposure and prioritize interventions (\Cref{sec:recommendations}).

\subsubsection{Likelihood.}
\label{sec:likelihood}

Two sets of factors may predict the likelihood that a service is flooded: agent capabilities and access, and existing features of the service.
 
\textit{Agent Capabilities and Access.}
Flooding overall becomes more likely as agent capabilities improve (Factor 1). Some capabilities, e.g. browser use, are likely particularly useful for government interaction (\cref{sec:ai-burdens}).

A service is more susceptible to flooding if the relevant agent capabilities are broadly accessible (Factor 2). Our cases (\cref{sec:exists}) are driven almost entirely by LLM text generation, a capability now freely available from multiple providers.
However, the most capable agents are paid services. Although inference is becoming cheaper overall, access costs or restrictions could prevent flooding, or restrict it to cases caused by wealthier users \citep{humlumUnequalAdoptionChatGPT2025, sharpAgenticInequality2025}.

\textit{Service Features.} The design of service interfaces can affect how readily certain capabilities translate into flooding (Factor 3) \citep{jordanRADARReadinessAI2026}. For example, even if an agent can collect all the relevant information and generate an application, it can have trouble interacting with portals that have complex login flows or JavaScript-heavy front-ends \citep{drouinWorkArenaHowCapable2024,gurRealworldWebAgentPlanning2023}. Similarly, digital agents cannot attend in-person appointments.

Services requiring more effort to submit could be more exposed to flooding (Factor 4). We would expect higher submission costs to suppress a higher share of true demand (\cref{sec:ai-burdens}). For example, voluntary tax returns can take significant effort, disincentivizing those who only expect a small return. As agents reduce these costs, it is therefore more likely the resulting demand increase causes strain. Conversely, hard limits on submission volume -- e.g. one annual tax return per citizen -- could limit how much demand agents can add. %

Finally, the expected benefit of submission likely predicts flooding itself, regardless of whether agents affect it (Factor 5). Falling submission costs alone can unlock latent demand \citep{currieTakeSocialBenefits2004}. For example, in [Case B] claimants more frequently respond to financially consequential claim denials because it is now easy, though there is no evidence their chance of success increases. We would expect this incentive to be stronger for services that offer higher financial reward. 

\subsubsection{Severity.}
\label{sec:evidence-severity}

Where flooding occurs, its impact could take two forms: first-order operational impacts from the demand surge itself, and second-order impacts and externalities from the responses governments adopt to address it.

\textit{Operational Impact.}
Operational impacts, such as backlogs, affect both the agency handling a service and its users. Their size depends on the efficiency and flexibility of the existing service (Factors 6--9), or its ``surge capacity'' \citep{hickRefiningSurgeCapacity2009, bonnettSurgeCapacityProposed2007}.
For public services, technical indicators of this capacity may include the use of structured and standardized data formats, which can enable ``straight-through'' processing of some cases without human intervention \citep{khannaStraightProcessingFinancial2008}, or digital identity systems, which can avoid time-consuming personhood confirmation \citep{naghmouchiPerspectivesNationalDigital2025}.

\textit{Response Impact.}
Beyond the direct impacts of flooding, the responses governments may choose often have trade-offs or negative externalities. We map these in \cref{sec:responses}. For example, closing a digital submission channel could restrict access to the service. These trade-offs depend both on what response is chosen, and how intensely it is implemented.

Legal constraints can affect which response governments choose. Rigid budget and entitlement policies (Factor 10) and legally mandated processing effort, such as response requirements \citep{kwokaSavingFreedomInformation2021} or right-to-explanation legislation \citep{buttaboniRegulatoryTaxonomyAI2026} (Factor 11), narrow the set of feasible responses.

More intense versions of responses have bigger externalities. For example, a more aggressive eligibility reduction for a welfare service affects more people. How strongly a government must intervene depends on the ``slack'' \citep{otooleDefenseBureaucracyPublic2010} in the budgets and rules around a service: how much demand could change before they would require updating \citep{hendrickRoleSlackLocal2006}. Slack is low where budgets assume take-up well below full entitlement (Factor 12), and where a successful submission has a high downstream cost, such as a benefit payout or a court proceeding (Factor 13).

\newenvironment{cit}
  {\begin{itemize}[leftmargin=*,nosep,topsep=0pt,partopsep=0pt,parsep=0pt]}
  {\end{itemize}}

\begin{table*}[t]
\centering
\renewcommand{\arraystretch}{1.2}
\begin{NiceTabularX}{\textwidth}{@{}c >{\raggedright\arraybackslash}p{2.4cm} L L L@{}}
\CodeBefore
  \rectanglecolor{red!18}{2-1}{3-1}
  \rectanglecolor{newblue!18}{4-1}{5-1}
\Body
\toprule
& \textbf{Approach} & \textbf{Sample Methods} & \textbf{Benefits} & \textbf{Drawbacks} \\
\midrule

\Block{2-1}{\hspace{\tabcolsep}\makebox[0.7cm]{\rotatebox[origin=c]{90}{\textbf{Suppress Demand}}}}
& \textit{Add Friction}
& \begin{cit}
    \item Fees
    \item In-person appointments
    \item Access limits, e.g. rate caps
    \item Closing digital channels
  \end{cit}
& \begin{cit}
    \item Fast to deploy
    \item Few prerequisites
    \item Proven to work
  \end{cit}
& \begin{cit}
    \item Can drive procedural inequality 
    \item Worsens user experience
    \item Some forms may erode as AI capabilities advance
  \end{cit} \\
\cmidrule(l){2-5}

& \textit{Reduce the Expected Benefit}
& \begin{cit}
    \item Reduce entitlement amounts
    \item Narrow eligibility criteria
    \item Fines for unsuccessful \newline applications
  \end{cit}
& \begin{cit}
    \item Robust to advances in agent capabilities
    \item Policy option usually \newline established
  \end{cit}
& \begin{cit}
    \item Legislative decision, can trade off other priorities
    \item Does not always address \newline operational impact
  \end{cit} \\
\midrule

\Block{2-1}{\hspace{\tabcolsep}\makebox[0.7cm]{\rotatebox[origin=c]{90}{\textbf{Increase Capacity}}}}
& \textit{Redesign the Service}
& \begin{cit}
    \item Structured interfaces with identity verification
    \item Deterministic processing pipelines
    \item Proactive ``push'' delivery
  \end{cit}
& \begin{cit}
    \item Technical architecture is established
    \item Can also improve citizen experience
  \end{cit}
& \begin{cit}
    \item High upfront investment
    \item Cross-agency coordination required
    \item Often needs legislative change; can take years
  \end{cit} \\
\cmidrule(l){2-5}

& \textit{Deploy Agents in Processing}
& \begin{cit}
    \item Intake and triage
    \item User correspondence
    \item Autonomous processing of routine cases
  \end{cit}
& \begin{cit}
    \item Scalable; likely effective
    \item Improves with capabilities improvements
  \end{cit}
& \begin{cit}
    \item Legal limits on automated decision-making
    \item Provider dependency and sovereignty risks
    \item Ethical and societal risks
  \end{cit} \\
\bottomrule
\end{NiceTabularX}
\caption{Overview of possible government responses to agentic flooding, grouped under two strategies.}
\label{tab:responses}
\end{table*}

\section{How Could Governments Respond?}
\label{sec:responses}

Governments explicitly respond to flooding in over half (56\%) of cases in our dataset. However, these actions are usually tightly scoped or non-binding, like releasing AI use guidance. Precedent suggests a much broader range of possible future responses, which we map to two categories (\Cref{tab:responses}): suppress demand for the affected services, or increase processing capacity to meet that demand. These responses would likely be effective, but each induces trade-offs. In particular, demand suppression restricts access to public services and can create procedural inequality.

\subsection{Suppress Demand}

Governments could suppress demand for services by making applications either more difficult or less attractive to complete, and have frequently done so in the past.

\paragraph{Add Friction.} Governments could (re-)raise the cost of submission until enough users are deterred from filing, thus suppressing quantitative flooding. Measures taken during previous demand surges include: financial barriers like fees \citep{kwokaSavingFreedomInformation2021}, procedural barriers like digital identity verification \citep{IDVerification2023}, and direct access limits like per-claimant rate caps \citep{rustVexatiousLitigantProblem2024}. An obvious candidate measure for agentic flooding is blocking bots from government websites, or restricting their permissions \citep{marroPermissionManifestsWeb2026}.

Most measures to introduce friction require little infrastructural change, have precedent, and can be deployed quickly, including when a surge is acute. In \statRespSuppressDemandN\ cases (\statRespSuppressDemandPct) in our dataset, governments have already responded with friction. For example, fees can be raised wherever a payment system is in place: Australia has considered reintroducing fees for FOI requests in response to a wave of AI-generated submissions [Case A]. Access can often be restricted with equally little lead time, as when Japanese authorities blocked submissions to a comment procedure by IP address [Case G]. Even where legal change is needed, it can be narrow: in [Case I], targeted reform stopped middle-man organizations from collecting fees on certain types of requests. %

However, introducing friction decreases the accessibility of government services. Friction disproportionately deters poorer, less digitally literate, and otherwise vulnerable users, effectively shaping who accesses public services \citep{peetersAdministrativeExclusionInfrastructurelevel2023}. Fees and in-person requirements can thereby create procedural inequality and, where legal services are concerned, undermine access to justice. Friction also worsens user experience of public services and, by extension, trust in government \citep{vandewallePublicServicePerformance2003,christensenTRUSTGOVERNMENTRelative2005}. Some measures, such as closing digital submission channels, could additionally prevent the accessibility improvements that AI agents promise \citep{yunImprovingCitizengovernmentInteractions2024}.

Friction-inducing measures may also face practical challenges. For one, some forms of friction could become less effective as agent capabilities improve, such as how CAPTCHAs no longer reliably identify human website visitors \citep{plesnerBreakingReCAPTCHAv22024}.
They may also be legally prohibited. For example, in many jurisdictions it is illegal to charge fees for applications to welfare services \citep{ssaStatePlansMedical2026}. 
Lastly, friction may not deter adversarial actors or directly address qualitative flooding (e.g. longer text submissions). 

\paragraph{Reduce the Expected Benefit.} Governments could also lower what a user stands to gain from submitting to a service. They could reduce entitlement amounts, narrow eligibility, or increase the cost of unsuccessful applications. Such revisions often reduce service demand, e.g. reforms to poverty assistance in the US in the 90s \citep{grogger2005welfare}. 

Entitlement revisions may be required regardless of whether they resolve flooding. Many services are budgeted assuming only limited take-up \citep{nicholsTargetingTransfersRestrictions1982}. For example, the UK Department for Work and Pensions explicitly anchors benefit budgets to historical take-up rates \citep{BenefitExpenditureCaseload2026}. Where budgets cannot be raised to meet full take-up, reducing benefits is a likely response.

A key challenge is that reducing benefits is not a purely executive decision, but a legislative decision. It may be harder to implement because of political priorities, and run counter to other aims, such as reducing poverty or helping the unemployed. %

\subsection{Increase Capacity}
\label{sec:increase-capacity}

Governments could also respond to flooding by increasing their processing capacity. Most obviously, they may choose to scale existing resources, e.g. by hiring more staff. However, tight budget constraints are common and existing processes are often inefficient \citep{dunleavyDigitalEraGovernance2006}.
We list two alternative options.

\paragraph{Redesign the Service.} Service redesign can include restructuring service interfaces, integrating identity verification, standardizing data formats, and providing some services proactively without applications \citep{dunleavyDesignPrinciplesEssentially2015}.
Such redesign could make government more efficient and free up resources for addressing flooding. Moreover, as discussed in \Cref{sec:risk-matrix}, structured forms and identity verification could decrease both the likelihood and the severity of flooding.

The main disadvantage is the long lead times and proactive financial investment required for implementation. 
Structural redesign requires cross-agency coordination and, in many cases, legislative change \citep{pollittPublicManagementReform2017, hoodPublicManagementAll1991}. Bigger projects can take years or even decades, such as introducing central registers or data standards; many Western governments have famously taken decades to get digital infrastructure in place despite well-documented benefits. As such, redesign is likely unavailable as a short-term measure. 

In \statRespRedesignServiceN\ cases (\statRespRedesignServicePct), we code the measures governments take in response to flooding as redesign -- though they are exclusively small changes, such as introducing a service to verify the legitimacy of case numbers [Case H], rather than broad structural reform.

\paragraph{Deploy Agents in Processing.} Governments could use AI agents themselves -- both to accelerate back-end processing, and to detect and prevent fraud. Governments are already taking first steps in this direction: in \statRespUseAiN\ cases (\statRespUseAiPct) in our dataset, they introduce AI tools explicitly in response to flooding, e.g. to analyze sentiment across large numbers of public comments [Case F] and to improve detection of AI-generated fake medical certificates [Case D]. In the vast majority of cases these are bounded tools handling one step in the process, consistent with established patterns of slow, piecemeal AI diffusion in public-sector organizations \citep{neumannExploringArtificialIntelligence2024}.

More comprehensive agent deployment lacks precedent, but evidence suggests it could scale processing throughput drastically. AI tools are already used in some countries for intake screening, triage, drafting of standard correspondence, and even autonomous handling of routine cases \citep{straubAIBureaucraticProductivity2024, oecdGoverningArtificialIntelligence2025}.
Government work is likely suitable for agent deployment because bureaucratic processes and rules are generally well-documented and exhaustive \cite{schmitzOversightStructuresAgentic2025}.
Benchmarks also suggest that agents are becoming increasingly competent in tasks similar to government work \citep{patwardhanGDPvalEvaluatingAI2025}.

However, there remain significant legal and practical risks regarding government use of agents. 
Laws often prevent automated government decision-making \citep{grimmelikhuijsenLegitimacyAlgorithmicDecisionmaking2022} and evaluating agents on government tasks remains difficult \citep{rystromAgentBenchmarksFail2026}.
Civil society groups regularly voice concerns about decision transparency, bias, and accountability diffusion \citep{AlgorithmicAccountabilityPublic2021}. 
Careless deployment could further expose government bodies to provider lock-in or sovereign-data risk \citep{jbatchikDigitalSovereigntyPractice2022}.

\section{Discussion and Recommendations}
\label{sec:discussion}

We find broad evidence of agentic flooding, but our work does not suggest it poses a severe \emph{operational} risk to governments. For one, responses to flooding so far are diverse and overall small in scale: piloting AI tools to help with individual steps, issuing guidance on AI use, or precise access limitations. Looking forward, while uncertainty remains about when particular agent capabilities arrive or diffuse widely, friction-inducing measures are likely to mitigate the operational impacts of flooding, as they have for previous demand surges following technological shifts \citep{verhoevenMyHowHave2024, ballaRespondingMassComputerGenerated2022}. %

However, adding friction worsens user experience of public services, potentially introduces procedural inequality, and prevents some of the accessibility gains that AI agents enable \citep{ilvesAgenticStateHow2025}. In contrast, capacity-building responses could address flooding without these trade-offs, but they require lead time, technical investment, and, in many cases, cross-departmental coordination; they are unlikely to be feasible once a surge is already underway.

One plausible trajectory is therefore that demand suppression again becomes a default response. Governments might reach for it because, when a surge occurs, it is the only intervention available on a short enough timeline. This pattern matches what \citet{lindblomScienceMuddling1959} terms ``muddling through'' -- iterative, short-term, and mostly reactive adaptation, which keeps government organizations operational, but does not usually optimize outcomes for the public.

\subsection{Three Near-Term Recommendations}
\label{sec:recommendations}

Preparing for flooding proactively may avoid governments having to (re)introduce friction to address it. We propose three near-term actions to this effect.

\paragraph{Auditing Exposure.} Governments should systematically audit their services for vulnerability to flooding. Key factors predicting the susceptibility of a service can be assessed today, and updated with future information about AI capabilities (\Cref{sec:risk-matrix}). A portfolio of services could be mapped and ranked against these factors, e.g. using our risk matrix, yielding a prioritized list of services and the response classes (\Cref{sec:responses}) most applicable to each. 

\paragraph{Developing a Digital Identity Strategy.} Strong identity verification could effectively mitigate quantitative flooding because it allows automated enforcement of per-claimant rate limits and makes adversarial flooding more difficult. 
Digital identity also supports pre-population of known data, thus lowering the effort required for entitled users to apply, and for governments to process their case. Over 100 jurisdictions already have digital identity infrastructure, but its maturity varies drastically \citep{metzID4DGlobalDataset2024}. Where feasible, integrating it into the most exposed services is likely to be a high-leverage action to preempt flooding.

\paragraph{Establishing Legal Certainty.} Several of the response options in \cref{sec:responses} are either illegal or of unclear legality in some jurisdictions. For example, many uses of AI in government processing are constrained by AI and administrative law \citep{buttaboniRegulatoryTaxonomyAI2026}, and it can be unclear where restricting free-form or digital submission channels, or raising fees, is lawful.
This ambiguity may itself be a barrier to preventative action.
Governments should commission internal legal reviews to clarify, for each response class, which forms of it are permissible under current law. This understanding would be valuable regardless of the order and intensity in which services are flooded.

\section{Limitations and Future Work}
\label{sec:limitations}

The methodology we employ aims to collect evidence of an emerging phenomenon across jurisdictions and domains. However, it allows neither statistical claims about prevalence, nor causal claims about AI's role in flooding. We suggest future research directions building on this work. To support such research, our dataset is published with this paper.

\paragraph{Collection Methodology.} 

Our semi-automated pipeline only collects positive instances of potential flooding and does not measure what share of public services (per jurisdiction) are assessed. As such, the statistics presented in \Cref{sec:exists}-\cref{sec:responses} cannot be taken to generalize within or beyond the studied countries, and we do not make comparative claims, e.g. of prevalence per country. 

Further sources of uncertainty or bias include the selection of countries and domains studied, the selection of ``candidate services'' presented during discovery, differing traditions of publicly discussing administrative matters, and the differing performance of LLMs in different languages (all prompts were given in English).

Further, while we increase our confidence in included cases by requiring explicit third-party attribution of the surge to AI, that criterion likely excludes many cases that meet our definition of flooding, and could wrongly include some others. We also do not require time-series volume data to include a case, and only \statSeriesCases\ possess it. Accordingly, we cannot make even correlative quantitative claims.

Finally, despite validating all cases with multiple instances of human review, we cannot exclude the risk of LLM hallucination. We only provide human-LLM agreement numbers, perform no inter-rater coding, and do not benchmark the performance of the LLM against humans.

Based on these limitations, a pressing direction for future work is more rigorous analysis of the relative prevalence of flooding for specific government services. Candidate methodologies may include quantitative analysis of volume time series, surveys of citizens about their use of AI agents, and analysis of agent usage data, where available. A further promising direction is more detailed analysis of the relevance of digital-government maturity for the risk and severity of flooding, e.g. analyzing how effective digital identity is at mitigation. 

\paragraph{Identification of Risk Factors and Mitigations.} 

Our contributions in \cref{sec:severity} and \cref{sec:responses} are largely based on precedent and theory, rather than our data.
We do not place our data in the risk matrix (\cref{sec:where-risk}). We also do not attempt to measure the success of any initial government responses we record, and the taxonomy of possible responses abstracts over jurisdiction-specific legal constraints.

This suggests a number of natural next steps for future work. 
The predictive value of the listed risk factors should be empirically validated within a single jurisdiction.
Government responses should be measured empirically for both their effectiveness and the externalities they produce.
Both the proposed risk matrix and response taxonomies should be continually updated as a result.

\section{Related Work}
\label{sec:related}

\paragraph{Concurrent Work.} \citet{piedrahitaAIPosesRisks2026} theorize ``congested bureaucracy'' as a result of AI-driven friction reduction; we validate and extend their framing here. \citet{shahAccessJusticeAge2026} find evidence that suggests AI use is driving a ``dramatic increase'' in cases before US federal courts.

\paragraph{Agents and Friction Reduction.} \citet{NBERc15309} identify that agents may lower search, communication, and contracting costs in markets. \citet{brynjolfssonAIsUseKnowledge2025} theorize that agents may reduce the cost of knowledge formalization.

\paragraph{AI and Government Interaction.} \citet{jordanRADARReadinessAI2026} measure the accessibility of government service interfaces to AI agents. \citet{yunImprovingCitizengovernmentInteractions2024} propose using LLMs to improve policy communication. \citet{majithiaCitizenQueryBenchmarkNovel2026} benchmark LLMs on answering citizen queries. \citet{zhangEnhancingCitizenGovernmentCommunication2025} find that AI tools improve some citizen-government interactions.

\paragraph{Administrative Burden.} \citet{madsenBurdensSludgeOrdeals2022} review how the transaction costs faced by citizens are studied under overlapping terms, including ``sludge'', ``red tape'', and ``ordeals''. \citet{zeckhauserStrategicSortingRole2019} examines a case where burdens are maintained intentionally.

\section{Conclusion}
\label{sec:conclusion}

\textit{Agentic flooding} is plausibly occurring across a broad range of government services, in the form of LLM-generated text submitted via permissive interfaces. Though it may intensify as agent capabilities improve, acute operational collapse appears unlikely. However, it \textit{is} likely that -- at least in some cases -- budgets will need adjusting or user experience will suffer, particularly where governments introduce friction to reduce service demand.

On a more positive note, mitigating agentic flooding is an opportunity to transform government services more broadly: many governments recognize the value of AI-enabled government interaction, and the digital-government literature has long advocated similar structural reforms for the benefit of the public \citep{dunleavyDigitalEraGovernance2006}.
While friction-based responses lock this potential further out of reach, proactive interventions to build capacity and redesign services could instead help to realize the user-experience benefits that have long been possible.

Governments can act now. Our analysis suggests three near-term, proactive measures whose value does not depend on tracking the progress of AI capabilities: auditing service vulnerability, developing digital-identity integration strategies for the most exposed services, and clarifying the legal basis for resilience-building measures.

\section*{Acknowledgements}

CS acknowledges funding by the Dieter Schwarz Foundation via the Hertie School. This work was partially completed during a seasonal fellowship at GovAI.

\appendix

\section{Research Scope} 
\label{app:scope}

\begin{table*}[!t]
\centering
\footnotesize
\setlength{\tabcolsep}{4pt}
\setlength{\aboverulesep}{0.9ex}
\setlength{\belowrulesep}{0.9ex}
\begin{tabular}{@{}r@{\hspace{5pt}}>{\RaggedRight\arraybackslash}p{0.16\textwidth} >{\RaggedRight\arraybackslash}p{0.27\textwidth} >{\RaggedRight\arraybackslash}p{0.26\textwidth} >{\RaggedRight\arraybackslash}p{0.22\textwidth}@{}}
\toprule
& \textbf{Surveyed domains} & \textbf{United Kingdom} & \textbf{United States} & \textbf{France} \\
 & & GOV.UK\newline \url{https://www.gov.uk/browse} &
USA.gov\newline \url{https://www.usa.gov/} &
Service-Public.fr\newline \url{https://www.service-public.gouv.fr/particuliers/vosdroits/theme} \\
\midrule
& \textit{Public Services} & & & \\
\addlinespace

\textbf{1} & Benefits \& Social Protection &
Benefits\newline Disabled people &
Government benefits\newline Disability services &
Social -- Santé \\
\arrayrulecolor{lightgray}\midrule[0.2pt]\arrayrulecolor{black}

\textbf{2} & Health Services &
-- &
Health &
Social -- Santé \\
\arrayrulecolor{lightgray}\midrule[0.2pt]\arrayrulecolor{black}

\textbf{3} & Citizenship, Identity \& Elections &
Citizenship and living in the UK\newline Births, deaths, marriages and care &
Voting and elections &
Papiers -- Citoyenneté -- Élections \\
\arrayrulecolor{lightgray}\midrule[0.2pt]\arrayrulecolor{black}

\textbf{4} & Education &
Education and learning\newline Childcare and parenting &
Education &
Famille -- Scolarité \\
\arrayrulecolor{lightgray}\midrule[0.2pt]\arrayrulecolor{black} 

\textbf{5} & Employment, Jobs \& Pensions &
Working, jobs and pensions\newline Employing people\newline Business and self-employed &
Jobs, labor laws and unemployment\newline Small business &
Travail -- Formation \\
\arrayrulecolor{lightgray}\midrule[0.2pt]\arrayrulecolor{black}

\textbf{6} & Housing \& Local Services &
Housing and local services &
Housing help &
Logement \\
\arrayrulecolor{lightgray}\midrule[0.2pt]\arrayrulecolor{black}

\textbf{7} & Immigration, Asylum \& Visas &
Visas and immigration\newline Passports, travel and living abroad &
Immigration and U.S. citizenship &
Étranger -- Europe \\
\arrayrulecolor{lightgray}\midrule[0.2pt]\arrayrulecolor{black}

\textbf{8} & Justice \& Legal \newline Services &
Crime, justice and the law &
Laws and legal issues\newline Scams and fraud &
Justice \\
\arrayrulecolor{lightgray}\midrule[0.2pt]\arrayrulecolor{black}

\textbf{9} & Tax \& Revenue \newline Administration &
Money and tax &
Taxes\newline Money and credit &
Argent -- Impôts -- Consommation \\
\arrayrulecolor{lightgray}\midrule[0.2pt]\arrayrulecolor{black}

\textbf{10} & Transport, Driving \& Licensing &
Driving and transport &
Travel &
Transports -- Mobilité \\

\midrule
& \textit{Other Channels} & & & \\
\addlinespace

\textbf{11} & Participatory \newline Processes & -- & -- & -- \\
\arrayrulecolor{lightgray}\midrule[0.2pt]\arrayrulecolor{black}

\textbf{12} & Regulatory Complaints \& Reporting &
-- &
Complaints &
-- \\
\arrayrulecolor{lightgray}\midrule[0.2pt]\arrayrulecolor{black}

\textbf{13} & FOI \& Transparency & -- & -- & -- \\

\midrule
\addlinespace

& \textit{Not Mapped}&
Environment and countryside &
Disasters and emergencies\newline Military and veterans\newline Innovation &
Associations\newline Loisirs -- Sports -- Culture \\

\bottomrule
\end{tabular}
\caption{Top-level citizen service categories on the national service portals of the UK, US, and France, and the domains we derive from them.}
\label{tab:portal-comparison}
\end{table*} 

\paragraph{Country selection.} We select twelve countries to 
capture variation geographically and along two institutional dimensions.
First, digital government maturity, as measured by the OECD Digital Government Index \citep{oecdGovernmentGlance20252025}. We include three of the five highest-rated countries: South Korea, Australia, and the United Kingdom. Three further included countries score above $0.8$, indicating middle-to-high maturity: Estonia, Denmark, and France. We include three that score under $0.8$, indicating lower maturity, e.g. higher reliance on non-digital processes: Brazil, Japan, and the Netherlands. Finally, we include three countries for which OECD data is not available: the United States, Germany, and Singapore. Beside aggregate maturity, these states vary in specific technical implementation of digital government. For example, four studied countries (Estonia, Denmark, Singapore, South Korea) have mandatory digital identity systems.

Second, administrative tradition \citep{painterTraditionPublicAdministration2010}: the list includes Anglo-American (UK, US, Australia), Napoleonic (France, Netherlands), Germanic (Germany), Scandinavian (Denmark), East Asian (Japan, South Korea, Singapore), Latin American (Brazil), and post-Soviet (Estonia) systems. Legally, these countries are split almost evenly between common-law and continental systems.

We select this list of countries to lower the chance of obvious confounders from these two dimensions, especially for our analyses of risk and prevalence of flooding. However, the list is evidently not representative, such that we do not make comparative or analytical claims about the distribution of flooding cases we find. Future work analyzing the relevance of these factors for flooding would likely be valuable, as we discuss in \cref{sec:limitations}.

\paragraph{Domain selection.} Within each country we scan a fixed list of government interaction channels. Existing taxonomies of the functions of government, e.g. COFOG by the \citet{oecdGovernmentGlance20252025}, are not suitable for this because they map government responsibilities, not citizen interfaces.

Instead, we compile a list of domains to scan by comparing the top-level
citizen service categories used by the national service portals of three of our sample countries: the UK, the US, and France (\cref{tab:portal-comparison}).
We map ten public service categories that return across all three of these portals. However, as clarified in our definition (\cref{sec:definition}),
some points of government interaction are not public services. We therefore append three domains for such channels: participatory processes, regulatory complaints and reporting, and transparency and access to information. Of these, only complaints have a top-level equivalent on any of the three portals.

This methodology produces a plausible list of 13 government domains, validated in three of the 12 studied countries and enabling the broad and indicative scans for flooding we seek to make. However, it is unlikely to be optimal. More detailed refinement of this domain list could improve it, e.g. by comparing against existing taxonomies or scanning more countries. Accordingly, we name this list of domains as a possible source of bias in \cref{sec:limitations}.
\section{Case Collection Pipeline}
\label{app:pipeline}

This appendix documents the pipeline used to collect the dataset presented in the paper. We describe the three pipeline stages and the prompts used
in each (\cref{app:stages}), the schema each case is coded
against (\cref{app:schema}), and the scaffolding that orchestrates the
pipeline (\cref{app:scaffold}). The full case records, prompts, data schema, and harness code are available in the linked dataset repository.

\subsection{Pipeline Stages}
\label{app:stages}

As depicted in \Cref{fig:pipeline-overview}, the pipeline has three stages: discovery, iteration, and finalization. An LLM is called during discovery and iteration, and all LLM calls return structured responses. We here describe each stage, provide summary statistics of cases processed in each, and give an overview of prompts and response schemas passed to the LLM.

\subsubsection{Discovery}
\label{app:discovery}

From a domain prompt combining a country and a domain, e.g. ``France, Health Services'', an LLM with web search access returns 20--30 candidate services to investigate. We survey 12 countries and the same 13 domains in each (\cref{app:scope}). Each candidate is returned as a stub, which is expanded into a case template deterministically. We then manually review the candidates and decide whether to investigate or remove each: of $2288$ candidates, we pass $2210$ (96\%) on to iteration.

\promptlabel{System prompt (opening)}

\begin{Verbatim}[breaklines=true,breaksymbolleft={},fontsize=\scriptsize,fontfamily=courier,xleftmargin=0pt]
You are a research assistant identifying candidate government services which may be experiencing "agentic flooding": surges of demand driven by AI use. You will receive a user input consisting of a country, and a domain of government services to investigate in that country. Your task is to return an initial list of 5-30 specific government services or interfaces which could be flooded in that domain. This is the first step in the pipeline, and these cases will be explored in detail afterward.

[...]
\end{Verbatim}

\promptlabel{User prompt}

\begin{Verbatim}[breaklines=true,breaksymbolleft={},fontsize=\scriptsize,fontfamily=courier,xleftmargin=0pt]
Country to analyze: {country} 
Government domain to analyze: {domain}.
\end{Verbatim}

\promptlabel{Response schema}

\begin{Verbatim}[breaklines=true,breaksymbolleft={},fontsize=\scriptsize,fontfamily=courier,xleftmargin=0pt]
Response: DiscoveryOutput

DiscoveryCandidate:
    service_name: str
    country: str
    government_body: str
    rationale: str

DiscoveryOutput:
    candidates: list[DiscoveryCandidate]
\end{Verbatim}

\subsubsection{Iteration}
\label{app:iteration}

Each selected case then cycles between a refinement call and a review call, with no human involvement, until the review call recommends finalization or removal, or a maximum of five cycles is reached.

\paragraph{Refinement.} The refinement call receives the current state of the case, a prompt to improve it by coding it against the case template \citep{dunivinScalingHermeneuticsGuide2025}, and any instructions from the previous iteration, review, or human reviewer. It returns a set of JSON Patch operations to apply to the case file, a recommendation to either continue iterating or pass the case to review, and a description of next steps for further iteration, if any. The harness applies the patches deterministically to the stored case and proceeds according to the recommendation.

The LLM does not produce source URLs as part of its response. Sources are extracted deterministically from
the web search grounding metadata returned with each API response, and added to the case data. For some elements of the case data, the model returns a text description of the source used to aid the human reviewer. We append any URL returned by API response metadata, regardless
of whether the model substantively used it. We manually verified all
cases listed in the final dataset, but the broader source list
in each case file often contains irrelevant links. We choose this approach as generating URLs as part of the response was very susceptible to hallucination \citep{huangSurveyHallucinationLarge2025}, while line-specific grounding is not available when using structured outputs.

\promptlabel{System prompt (opening)}

\begin{Verbatim}[breaklines=true,breaksymbolleft={},fontsize=\scriptsize,fontfamily=courier,xleftmargin=0pt]
You are a research assistant helping build a dataset of "agentic flooding" cases for an academic paper. Agentic flooding is when AI use by citizens drives a surge in demand on a government service, either in volume, in complexity, or both. The paper asks whether this is happening, where, and what governments could do about it.

Your job is to improve a single case record by performing web search to fill in or correct fields, and to judge whether the case meets each of three inclusion criteria. 

[...]
\end{Verbatim}

\promptlabel{User prompt}

\begin{Verbatim}[breaklines=true,breaksymbolleft={},fontsize=\scriptsize,fontfamily=courier,xleftmargin=0pt]
{Case JSON}
\end{Verbatim}

\promptlabel{Response schema}

\begin{Verbatim}[breaklines=true,breaksymbolleft={},fontsize=\scriptsize,fontfamily=courier,xleftmargin=0pt]
Response: IterationOutput

Patch:
    path: str
    value: Any
    rationale: str

IterationOutput:
    patches: list[Patch]
    summary_of_changes: str
    next_steps: str
    recommendation: str
\end{Verbatim}

\begin{table}[tp]
\centering
\scriptsize
\begin{tabular}{@{}p{0.68\columnwidth}l@{}}
\toprule
\textbf{Field} & \textbf{Type} \\
\midrule
\multicolumn{2}{@{}l}{\textit{Identification}} \\
\path{case_id}             & string \\
\path{service_name}        & string \\
\path{country}             & string \\
\path{jurisdiction_detail} & string \\
\path{government_body}     & string \\
\path{include}             & bool \\
\path{tldr}                & string \\
\path{analysis}            & string \\

\addlinespace
\multicolumn{2}{@{}l}{\textit{Types of Flooding} (\cref{sec:variations})} \\
\path{flooding.qualitative}  & bool \\
\path{flooding.quantitative} & bool \\

\addlinespace
\multicolumn{2}{@{}l}{\textit{Mechanism} (inclusion criterion 1)} \\
\path{mechanism.meets_inclusion_criteria}   & bool \\
\path{mechanism.inclusion_explanation}      & string \\
\path{mechanism.ai_description}             & string \\
\path{mechanism.primary_is_text_generation} & bool \\
\path{mechanism.interface_description}      & string \\

\addlinespace
\multicolumn{2}{@{}l}{\textit{Evidence of change} (inclusion criterion 2)} \\
\path{evidence.meets_inclusion_criteria} & bool \\
\path{evidence.explanation}              & string \\

\addlinespace
\multicolumn{2}{@{}l}{\textit{AI attribution} (inclusion criterion 3)} \\
\path{attribution.meets_inclusion_criteria} & bool \\
\path{attribution.explanation}              & string \\
\path{attribution.source}                   & string \\
\path{attribution.description}              & string \\

\addlinespace
\multicolumn{2}{@{}l}{\textit{Volume time series}} \\
\path{volume.unit}                      & string \\
\path{volume.annual.{2018..2025}}       & number \\
\path{volume.source_notes.{2018..2025}} & string \\

\addlinespace
\multicolumn{2}{@{}l}{\textit{Government response} (\cref{sec:responses})} \\
\path{response.action_explicitly_for_ai} & bool \\
\path{response.action_for_general_surge} & bool \\
\path{response.description}              & string \\
\path{response.categories}               & string[] \\

\addlinespace
\multicolumn{2}{@{}l}{\textit{Sources}} \\
\path{sources[].url}    & string \\
\path{sources[].origin} & string \\

\addlinespace
\multicolumn{2}{@{}l}{\textit{Pipeline metadata}} \\
\path{pipeline.status}                  & string \\
\path{pipeline.discovery_country}       & string \\
\path{pipeline.discovery_domain}        & string \\
\path{pipeline.discovery_rationale}     & string \\
\path{pipeline.iteration_count}         & integer \\
\path{pipeline.last_summary_of_changes} & string \\
\path{pipeline.last_next_steps}         & string \\
\path{pipeline.last_review_concerns}    & string \\
\path{pipeline.human_review_comment}    & string \\
\path{pipeline.last_error}              & string \\
\bottomrule
\end{tabular}
\caption{Fields collected for each case, grouped by section. Field descriptions are released with the dataset repository.}
\label{tab:schema}
\end{table}

\paragraph{Review.} The review call skeptically evaluates the current case and returns a recommendation of \texttt{finalize}, \texttt{remove}, or \texttt{iterate}, with optional field-level concerns. It does not modify the case. Its output becomes input either to a human decision, when it recommends finalization or removal, or to the next refinement call, when it recommends further iteration.

\promptlabel{System prompt (opening)}

\begin{Verbatim}[breaklines=true,breaksymbolleft={},fontsize=\scriptsize,fontfamily=courier,xleftmargin=0pt]
You are a skeptical reviewer evaluating a candidate case for an academic dataset of "agentic flooding" — situations where AI use by citizens drives surges in demand on government services. A research assistant has iterated on this case across multiple passes. Your job is to decide what happens to it next: finalize it into the dataset, send it back for more work, or remove it as unsuitable.

[...]
\end{Verbatim}

\promptlabel{User prompt}

\begin{Verbatim}[breaklines=true,breaksymbolleft={},fontsize=\scriptsize,fontfamily=courier,xleftmargin=0pt]
{Case JSON}
\end{Verbatim}

\promptlabel{Response schema}

\begin{Verbatim}[breaklines=true,breaksymbolleft={},fontsize=\scriptsize,fontfamily=courier,xleftmargin=0pt]
Response: ReviewOutput

ReviewConcern:
    field_path: str
    concern: str

ReviewOutput:
    recommendation: str
    rationale: str
    specific_concerns: list[ReviewConcern]
\end{Verbatim}

\subsubsection{Finalization}
\label{app:finalization}

We manually review every case the review call recommends for finalization or removal. Each is either added to the dataset, removed, or returned to iteration with concrete improvement instructions. Of $2210$ cases, we take the recommended action in $2182$ (98\%); the most recommended action is removal (95\% of cases).

\subsection{Case Schema}
\label{app:schema}

Each case is stored as a JSON document. \Cref{tab:schema} lists the fields
collected and their types, grouped by section. The full schema, including a description of every field, is available in the dataset repository. The schema
mirrors the structure of our definition
(\cref{sec:service-flooding}). The three sections corresponding to our inclusion criteria (\cref{sec:case-selection}) -- \texttt{mechanism}, \texttt{evidence},
\texttt{attribution} -- each contain a \texttt{meets\_inclusion\_criteria} boolean,
and a case is included in the dataset only if all three are true.

\subsection{Scaffolding}
\label{app:scaffold}

\paragraph{Storage.} Each case is stored as a JSON file on disk, named by its
\texttt{case\_id}, and versioned in the repository. An append-only log records the output of every iteration, including the patches applied, the LLM's summary of changes, and the recommendation
returned.

\paragraph{Human-in-the-loop.} A minimal web interface allows a human
reviewer (the first author) to make two decisions: (i) which candidates from
discovery to advance to iteration, and (ii) for each case returned by the
review stage, whether to finalize, exclude, or return to iteration with
optional written feedback.

\paragraph{Models, cost, and reproducibility.} 
All \textit{discovery} and \textit{review} calls used \texttt{Gemini-3.1-Pro-Preview} via the Gemini API; all \textit{iteration} calls used \texttt{Gemini-3.1-Flash-Lite}.
Web search grounding was enabled for \textit{discovery} and \textit{iteration}, but not for \textit{review} calls. Temperature was set to 0 for all calls.
Total token cost for producing the final dataset was approximately
\texttt{\$200}. Full prompts and orchestration code are included in the linked dataset repository to support replication. Both the models used and the web search grounding component cannot be guaranteed to be deterministic; re-running the pipeline may not produce an identical dataset.

\bibliography{flooding}

\end{document}